# ALPHA-ROOTING COLOR IMAGE ENHANCEMENT METHOD BY TWO-SIDE 2-D QUATERNION DISCRETE FOURIER TRANSFORM FOLLOWED BY SPATIAL TRANSFORMATION

Artyom M. Grigoryan[1], Aparna John[1], Sos S. Agaian[2]

[1]University of Texas at San Antonio, San Antonio, TX, USA 78249
[2]City University of New York / CSI

## ABSTRACT

*In this paper a quaternion approach of enhancement method is proposed in which color in the image is considered as a single entity. This new method is referred as the alpha-rooting method of color image enhancement by the two-dimensional quaternion discrete Fourier transform (2-D QDFT) followed by a spatial transformation. The results of the proposed color image enhancement method are compared with its counterpart channel-by-channel enhancement algorithm by the 2-D DFT. The image enhancements are quantified to the enhancement measure that is based on visual perception referred as the color enhancement measure estimation (CEME). The preliminary experiment results show that the quaternion approach of image enhancement is an effective color image enhancement technique.*

## 1. INTRODUCTION

The color in an image is resolved to 3 or 4 color components and the images corresponding to these color components are stored in 3 or 4 separate channels. Therefore, the original color of the color image is only obtained by summing color components. In most of the popular enhancement algorithms applied onto color images, the image is first separated into color component images and then the given algorithm is applied to each of the component images. Such approach for color image enhancement does not give the complete effect of the algorithm on the color of images. All enhancement algorithms on color images should be applied in such a way that color of the image is taken as a single entity. In Ell and Sanguine's works[11]-[18], the color image is represented in the quaternion space. The quaternion number is a four-dimensional hyper-complex number. Representation of color image as a quaternion image enables us to represent each of the three color components as a vector in orthogonal color-cube space. The quaternion representation of the color image helps in processing the image as the sum of all color component images. That is, the color is taken as a single entity and the color image is considered as a single unit.

In this paper, we develop a color image enhancement method, that is based on the quaternion approach of color image enhancement in the frequency domain. The proposed method is referred to as the alpha-rooting method of color image enhancement by the two-side two-dimensional quaternion discrete Fourier transform (2-D QDFT) followed by a spatial transformation. Alpha-rooting based methods are frequency domain image enhancement techniques and effectively used on both grayscale and color images when processed channel-by-channel. The alpha-rooting method in the quaternion space is also showing good enhancement results. The results of enhancement by the proposed method is compared with the channel-by-channel color image enhancement technique with the same method but based on the 2-D DFT. Enhancements in color images are measured quantitatively by the enhancement measure suggested by the authors Grigoryan and Agaian. This particular enhancement measure, referred to as the color enhancement measure estimation (CEME), is based on the visual perception of images. The alpha value that gives the maximum measure value is chosen as the optimum alpha for the proposed alpha-rooting method.

## 2. COLOR IMAGE ENHANCEMENT METHODOLOGY

### 2.1. Enhancement Algorithm Using the 2-D QDFT

#### 2.1.1. Quaternion Algebra

Quaternion numbers are four-dimensional hyper-complex numbers [3],[9]-[18] that are represented in Cartesian form as

$$q = a + ib + jc + kd. \qquad (1)$$

That is, every quaternion number $q$ has a scalar part, $S(q)=a$ and a vector part, $V(q)=ib+jc+kd$. The imaginary units i, j, and k are related as

$$i^2 = j^2 = k^2 = ijk = -1, \qquad (2)$$

$$ij = -ji = k; jk = -kj = i; ki = -ik = j.$$

The basis {1, i, j, k} is the most common and classical basis to express quaternions. There are many different choices for $q$. Given two pure unit quaternions μ and ψ, that is, μ²=ψ²= −1 that are orthogonal to each other, that is, μ ⊥ ψ, the set {1, μ, ψ, μψ} is a basis for any quaternion $q$.

The magnitude of the quaternion is defined as

$$|q| = \sqrt{\|q\|} = \sqrt{a^2 + b^2 + c^2 + d^2}, \qquad (3)$$

where ‖$q$‖ is the norm of the quaternion. When ‖$q$‖ =1, $q$ is said to be unit quaternion. When the scalar part, $S(q)=0$, then that quaternion is referred to as a pure quaternion.

An important property in the quaternion algebra is the fact that the product of two quaternions may be non-commutative. That is, for a given two quaternion numbers $p$ and $q$.

$$pq = qp \text{ or } pq \neq qp. \tag{4}$$

### 2.1.2. Color image as a quaternion number

The color image can be represented in the quaternion space[9],[11]-[16],[34]. Depending on the color model, the color image have three or four channels. In the case of three channel color models like the RGB, or XYZ, the color images can be represented as pure quaternions. For example, the color image f (n,m) in the RGB color model, when represented as quaternion numbers, takes the form

$$f(n,m) = i\, r(n,m) + j\, g(n,m) + k\, b(n,m) \tag{5}$$

where r(n,m), g(n,m) and b(n,m) are the red, green, and blue components, respectively

When a color image is represented as quaternions, we get a correlation between all color components. In the usual color image enhancements, this correlation between the color components is not obtained because of the enhancement algorithm that is applied to each channel separately. The real component, $a$ of the quaternion image can be chosen in many different ways[6].

### 2.1.3. Two-sided 2-D Quaternion Fourier Transform

The discrete Fourier transform of 2-D hyper-complex numbers is referred as 2-D QDFT. As mentioned before, the product of two quaternions is non-commutative, in general. Thus, based on the position of the kernel of the transform with respect to the signal there are many versions of 2-D QDFT. In the two-sided 2-D QDFT, the signal is sandwiched by the two transform kernels

$$W_j^t = \exp\left(-\frac{j2\pi t}{N}\right) = \cos\left(\frac{2\pi t}{N}\right) - j\sin\left(\frac{2\pi t}{N}\right), t = 0: (N-1),$$

$$W_k^t = \exp\left(-\frac{j2\pi t}{M}\right) = \cos\left(\frac{2\pi t}{M}\right) - k\sin\left(\frac{2\pi t}{M}\right), t = 0: (M-1),$$

and is defined as

$$\text{QDFT}, F(p,s) = \sum_{n=0}^{N-1} W_j^{np} \left[\sum_{m=0}^{M-1} f_{n,m} W_k^{ms}\right], \tag{6}$$

$$p = 0: (N-1), s = 0: (M-1).$$

The inverse transform is calculated by

$$\text{IQDFT}, f(n,m) = \frac{1}{NM} \sum_{p=0}^{N-1} W_j^{-np} \left[ \sum_{s=0}^{M-1} F_{p,s} W_k^{-ms} \right], \qquad (7)$$

$$n = 0:(N-1), m = 0:(M-1),$$

Similarly, in the left-sided 2-D QDFT the position of kernels of the transform is on the left side of the signal and in the right-sided 2-D QDFT, the position of the kernels of the transform is on the right side. Since the product of two quaternion numbers is non-commutative, the transform obtained by each definition of the 2-D QDFT is different [27].

### 2.1.4. Alpha-rooting

In the alpha-rooting method of image enhancement [2]-[4] [17] –[26],[28], for each frequency point (p,s), the magnitude of the quaternion discrete Fourier transform are transformed as

$$|F_{p,s}| \rightarrow |F_{p,s}|^\alpha \qquad (8)$$

The choice of the $\alpha$ lies between (0,1) [4],[17]-[20].

### 2.1.5. Color Enhancement Measure Estimation (CEME)

The color enhancement measure estimation (CEME) is an enhancement measure [4]-[8],[35] based on the contrast of the images. The proposed enhancement measure is a modification of a theory on contrast perception. Weber's law explains that the contrast perception is constant at high values of luminance and low spatial frequency, and Hunt's effect and Steven's effect mention that the contrast and color saturation increases with luminance. To calculate the CEME value, the 2-D discrete image of the size $N \times M$ is divided[21] by $k_1 k_2$ blocks of size $L_1 \times L_2$ each, where $kn = \lfloor N_n/L_n \rfloor$, for $n=1,2$. For the RGB color model, when the image is transformed by the proposed enhancement algorithm,

$$T_\alpha: f = (f_R, f_G, f_B) \rightarrow \hat{f} = (\hat{f}_e, \hat{f}_R, \hat{f}_G, \hat{f}_B), \qquad (9)$$

where $\hat{f}_e$ is referring to the scalar component of the quaternion image, obtained after taking the transform, the CEME value is calculated by

$$E_q(\alpha) = CEME_\alpha(\hat{f}) = \frac{1}{k_1 k_2} \sum_{k=1}^{k_1} \sum_{l=1}^{k_2} 20 \log_{10} \left[ \frac{MAX_{k,l}(\hat{f}_e, \hat{f}_R, \hat{f}_G, \hat{f}_B)}{MIN_{k,l}(\hat{f}_e, \hat{f}_R, \hat{f}_G, \hat{f}_B)} \right]. \qquad (10)$$

## 2.2. Enhancement Method with the 2-D DFT

The processing of the image can be done on each channel separately. The 2-D DFT and the subsequently the alpha-rooting can be done on each channel individually. The enhancement of the image by the alpha-rooting causes a modification in the frequency domain. But, further enhancement of the image by spatial transformation techniques, such as the histogram equalization, gives an additional enhancement in the spatial domain.

### 2.2.1 The Enhancement Measure Estimation (EME) for Grayscale Images

The enhancement measure estimation (EME) for grayscale images is an enhancement measure [4]-[8] based on the contrast of the images. To calculate the EME value, the 2-D discrete image of size N×M is divided[21] by $k_1 k_2$ blocks of size $L_1 \times L_2$ each, where $k_n = \lfloor N_n / L_n \rfloor$, for $n=1,2$. When each channel in an image is transformed by the enhancement algorithm,

$$T_\alpha : f \rightarrow \hat{f} \qquad (11)$$

where $\hat{f}$ is referring to the enhanced image, the EME value is calculated by

$$E_q(\alpha) = EME_\alpha|(\hat{f}) = \frac{1}{k_1 k_2} \sum_{k=1}^{k_1} \sum_{l=1}^{k_2} 20\, log_{10} \left[ \frac{\text{MAX}_{k,l}(\hat{f})}{\text{MIN}_{k,l}(\hat{f})} \right]. \qquad (12)$$

In the channel-by-channel approach of color image enhancement algorithms, the enhancement measure is determined for images in each channel.

### 2.3. Spatial Transformation by Histogram Equalization

The enhanced images by the alpha-rooting method with the 2-D QDFT and the alpha-rooting method by 2-D DFT are further enhanced by spatial transformations. The spatial transformation by the histogram equalization of color images is applied onto the enhanced images by the alpha-rooting method by both 2-D QDFT and 2-D DFT. In color images, the histogram equalization is done by converting the color image in the RGB to models like HSV, YUV, YCbCr or any other similar color models. For example, in the case of HSV model, the V component is histogram equalized and then the modified HSV image is converted back to get the RGB image. In the RGB color model, the intensity of the image is spread as intensities of the red, green and blue colors. Therefore, the separating of each color channel and taking the histogram equalization of each channel does not work right. The intensities of the color image are stored as 24 bits which are a combination of the red, green, and blue color.

## 3. EXPERIMENT RESULTS

Figure 1 shows a comparative study of the enhanced images by the alpha-rooting method both by the 2-D QDFT and by the 2-D DFT. The original image "tree_color.tiff" in part (a) is enhanced by the alpha-rooting method by the 2-D QDFT and then further enhanced by the histogram equalization of the resulting image.

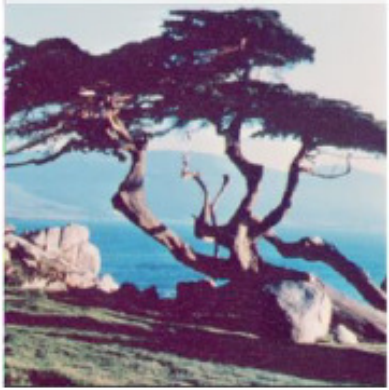

(a)

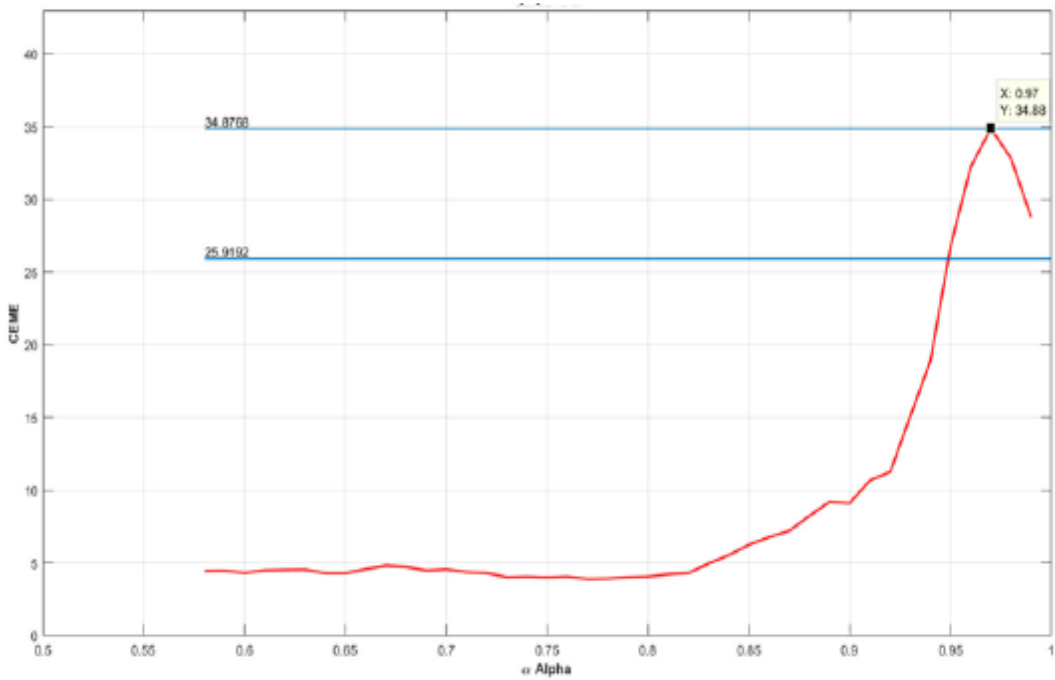

(b)

Figure 1: (a) Original Image "tree_color.tiff"[29]; (b) Plot of CEME vs alpha for alpha-rooting method by the 2-D QDFT.

The resulting images are as shown in Fig. 1(c) and (d) respectively. The choice of alpha value for the alpha-rooting method by the 2-D QDFT is from the plot of CEME vs alpha, as shown in Fig. 1 (b). The alpha that gives the maximum CEME measure is chosen for the image processing.

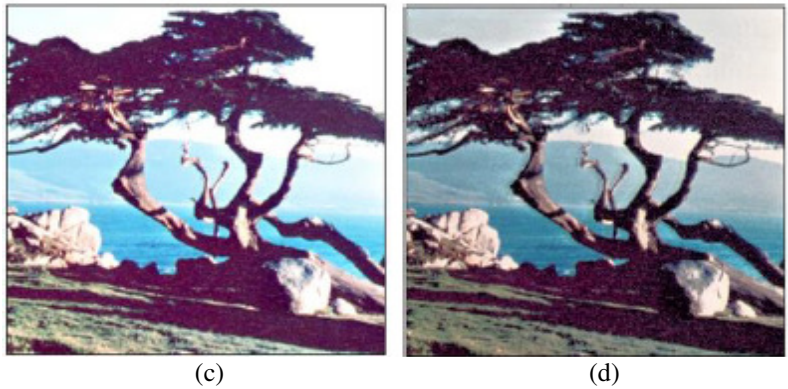

(c)                                                                   (d)

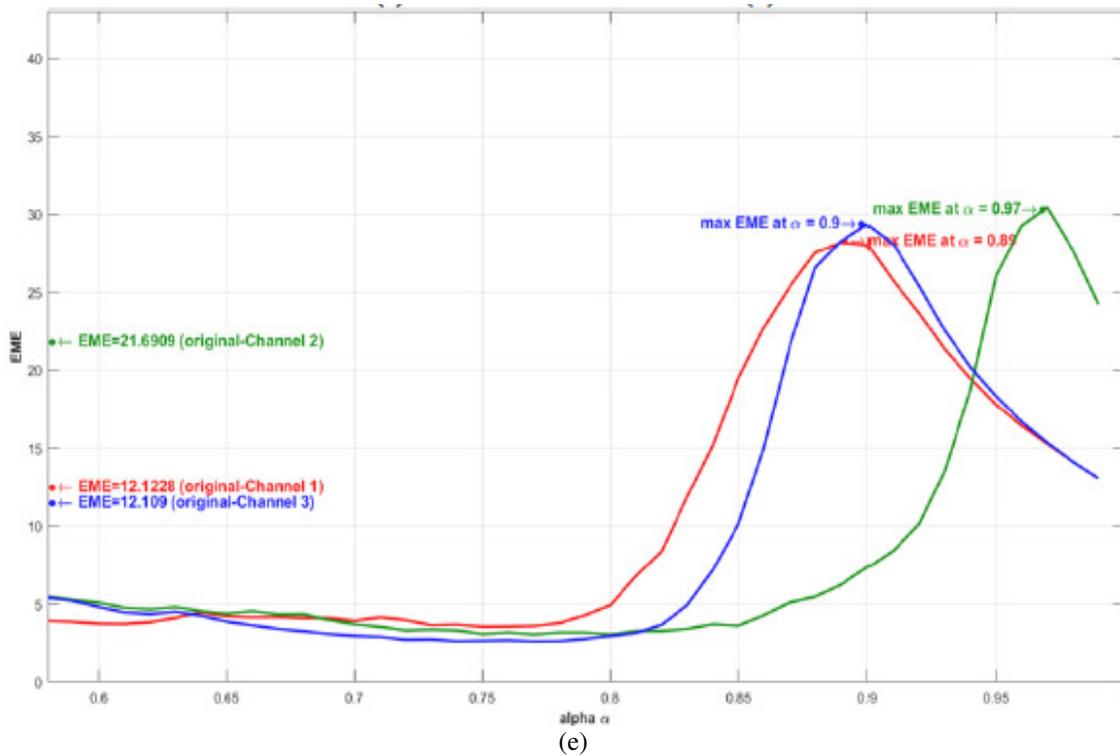

(e)

Figure 1: (continued) (c) Enhanced image by alpha-rooting by 2-D QDFT; (d) Histogram equalised image of the image in (c); (e) Plot of EME vs alpha for alpha-rooting by 2-D DFT for each channel (R,G,B).

The same "tree_color.tiff" image is also enhanced by the alpha-rooting by the 2-D DFT, when the enhancement is done channel-by-channel. The alpha required for each channel is obtained by the plot of EME vs alpha for each channel, that is, for the red, green and blue channels. The plot of EME vs alpha for each channel for "tree_color.tiff" is shown in Fig. 1(e). The enhanced images by the alpha-rooting by the 2-D DFT are shown in Fig. 1(f) and the spatial transformation by the histogram equalization of the image in 1(f) is shown in Fig. 1(g). Table 1 gives a comparative study of the two different algorithms of the alpha-rooting methods by the 2-D QDFT and by the 2-D DFT. It can be seen from Table 1 that the CEME value of the enhanced image by the alpha-rooting by the 2-D QDFT and by the 2-D DFT is almost of the same value with the CEME value of the latter method giving a slightly higher value. But when both resulting images are further enhanced by the histogram equalization, the CEME value shows a much higher value for the image enhanced prior the alpha-rooting by the 2-D QDFT. Since in both cases the images are in the RGB color model, the RGB image is first converted to the HSV color model and the V component is histogram equalized. The histogram equalized HSV image is converted back to the RGB image. The resulting enhanced images are shown in parts (d) and (g) of Fig. 1.

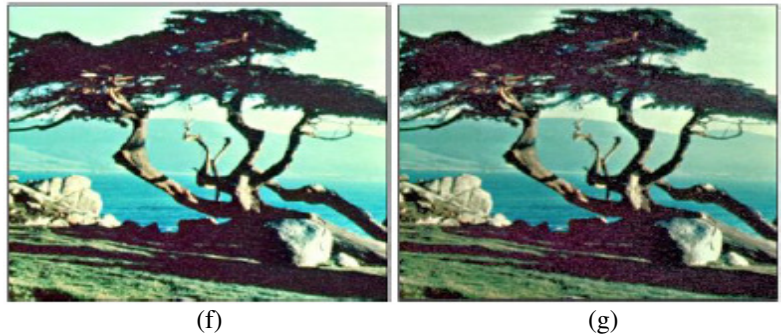

(f)            (g)

Figure 1: (continued) (f) Enhanced image by alpha-rooting by the 2-D DFT; (g) Histogram equalized image of image in (f).

Table 1: Alpha and CEME/EME values of Original and Enhanced Image of "tree_color.tiff".

| "tree_color.tiff" | CEME | Alpha | EME |
|---|---|---|---|
| Original Image | 25.9192 |  | R: 12.1228 |
|  |  |  | G: 21.6909 |
|  |  |  | B: 12.1090 |
| 2-D QDFT Alpha-Rooting | 34.8768 | 0.97 | - |
| 2-D QDFT Alpha-Rooting With Spatial Transformation | 49.0341 | 0.97 | - |
| 2-D DFT Alpha-Rooting | 35.8855 | R: 0.89 | 28.1572 |
|  |  | G: 0.97 | 30.4253 |
|  |  | B: 0.90 | 29.3055 |
| 2-D DFT Alpha-Rooting with Spatial Transformation | 40.7045 | R: 0.89 |  |
|  |  | G: 0.97 |  |
|  |  | B: 0.90 |  |

Figure 2 shows the enhancement of the original color image "image19.jpg" shown in Fig. 2(a). The alpha value for the image enhancement algorithm is obtained from the plot of CEME vs alpha in Fig. 2(b). The alpha which gives the maximum CEME value is chosen for the alpha in the alpha-rooting method by 2-D QDFT. The image enhanced and its further spatial transformation enhancement by histogram equalization is shown in parts (c) and (d) respectively of Fig. 2. For the enhancement of the image "image19.jpg" by the alpha-rooting method by the 2-D DFT, three different values of alpha are chosen from the plot in Fig. 2(e). The alpha-rooting method by the 2-D DFT is done separately on each channel with the best alpha value of the corresponding channel. The enhanced images by the method of alpha-rooting by the 2-D DFT and its histogram equalized images are shown respectively in parts (f) and (g) of Fig. 2.

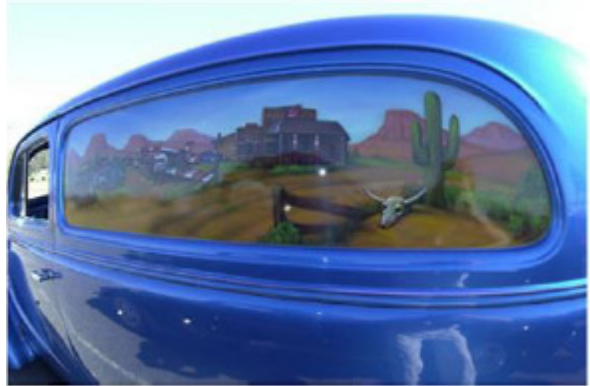

(a)

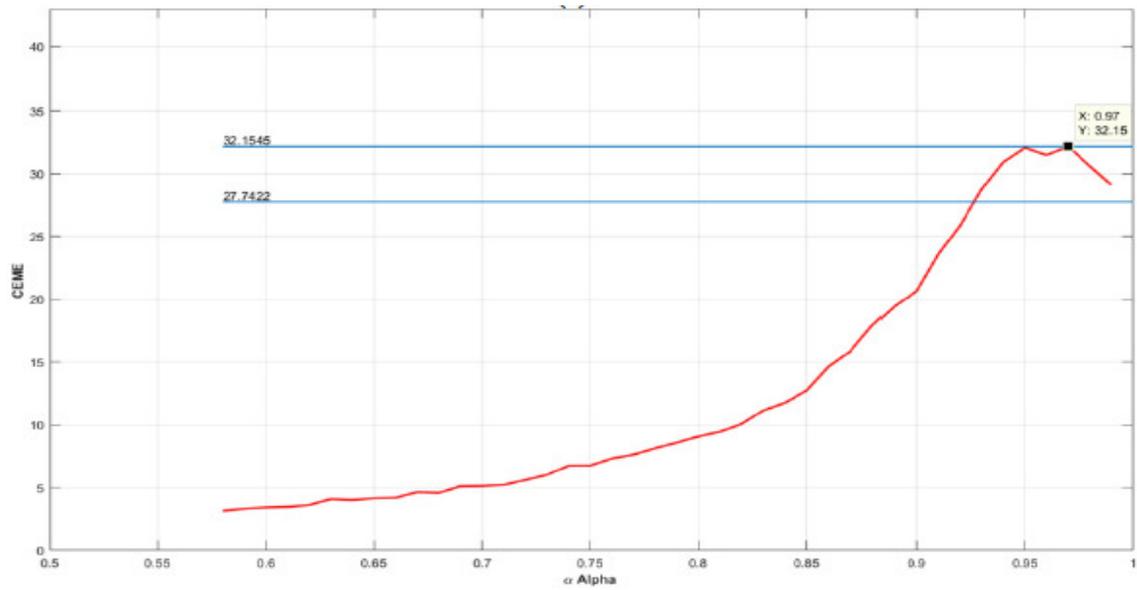

(b)

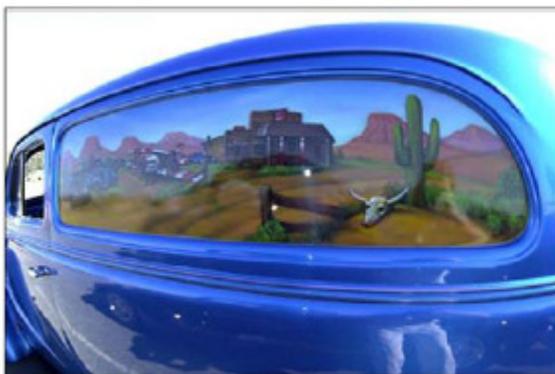 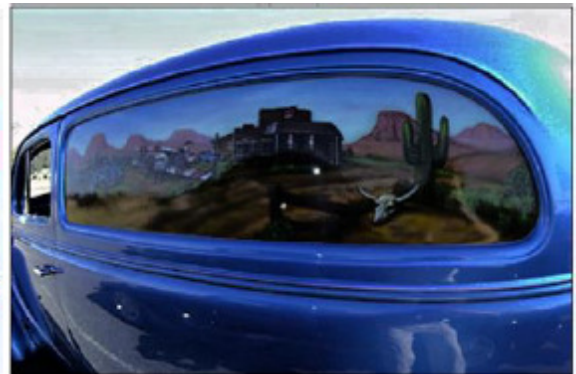

(c)            (d)

Figure 2: (a) The original Image "image19.jpg"[30]; (b) The plot of CEME vs alpha for alpha-rooting method by the 2-D QDFT; (c) Enhanced image by alpha-rooting by the 2-D QDFT; (d) Histogram equalised image of the image in (c).

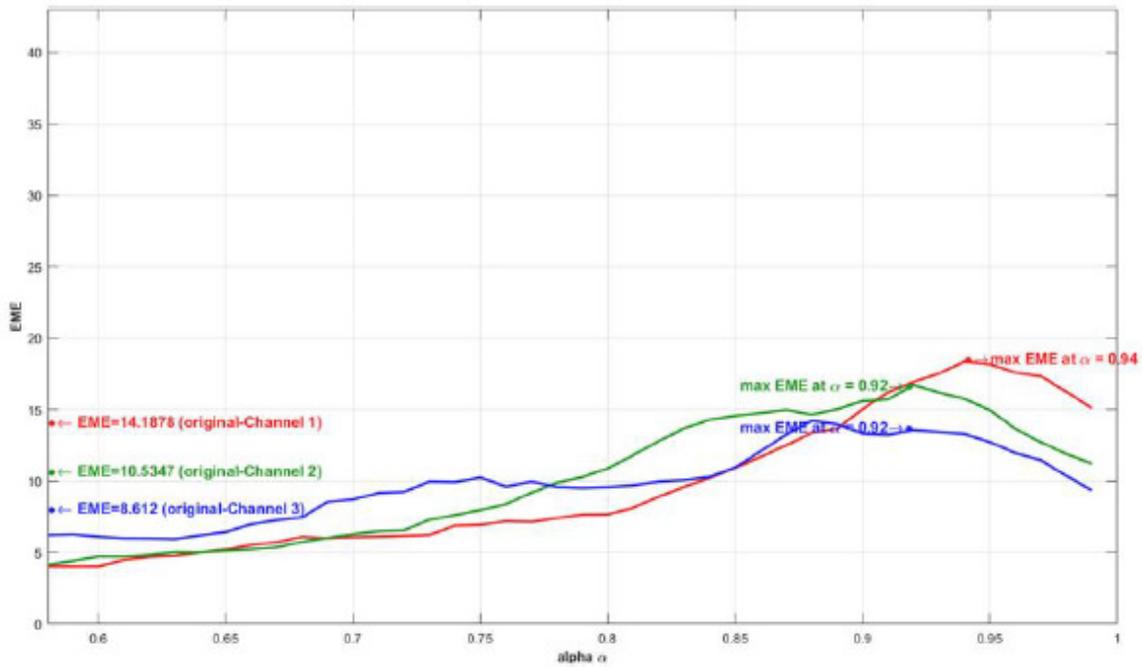

(e)

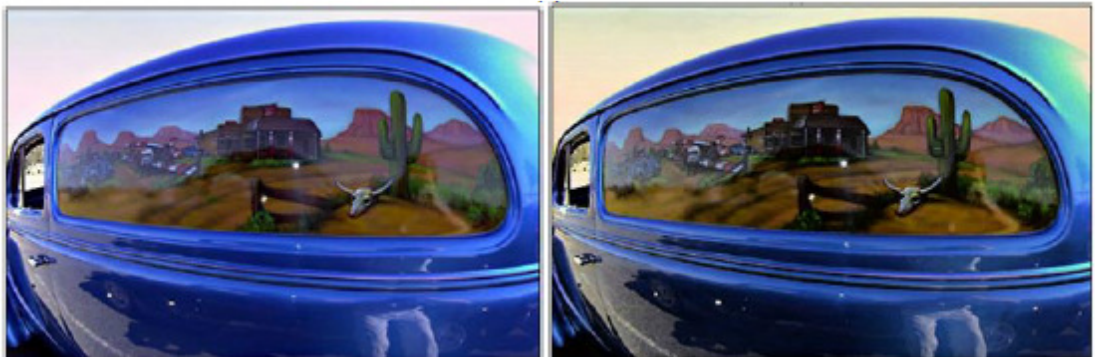

(f)                                                   (g)

Figure 2: (continued) (e) Plot of EME vs alpha for the alpha-rooting by the 2-D DFT for each channel (R,G,B), (f) Enhanced image by alpha-rooting by the 2-D DFT; (g) Histogram equalised image of image in (f).

Table 2 gives a comparative study of the enhancement by both methods of alpha-rooting, one by the 2-D QDFT and the other by the 2-D DFT. We can see from Table 2, that the enhancement measure CEME value of the enhanced image by the method of the 2-D QDFT is higher than the channel-by-channel approach by the 2-D DFT. And also, for the further spatial transformation by histogram equalization, the image prior enhanced by the 2-D QDFT gives a higher CEME value than the image prior enhanced by the 2-D DFT. The best alpha value chosen for alpha-rooting by the 2-D QDFT is 0.97 and for the method by the 2-D DFT the three respective best alpha values for red, green and blue channels are 0.94, 0.92, and 0.92.

Table 2: Alpha and CEME/EME values of Original and Enhanced Image of "image19.jpg"

| "image19.jpg" | CEME | Alpha | EME |
|---|---|---|---|
| Original Image | 27.7422 | | R: 14.1878 |
| | | | G: 10.5347 |
| | | | B: 8.6120 |
| 2-D QDFT Alpha-Rooting | 32.1545 | 0.97 | - |
| 2-D QDFT Alpha-Rooting With Spatial Transformation | 34.6855 | 0.97 | - |
| 2-D DFT Alpha-Rooting | 29.6856 | R: 0.94 | 18.3581 |
| | | G: 0.92 | 16.7192 |
| | | B: 0.92 | 13.5493 |
| 2-D DFT Alpha-Rooting with Spatial Transformation | 39.5148 | R: 0.94 | |
| | | G: 0.92 | |
| | | B: 0.92 | |

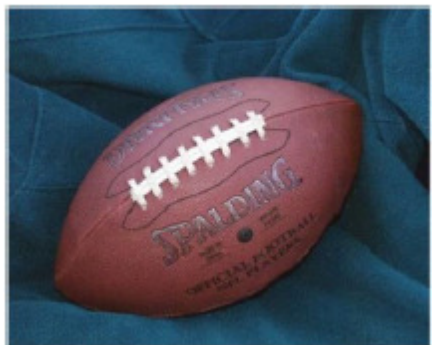

(a)

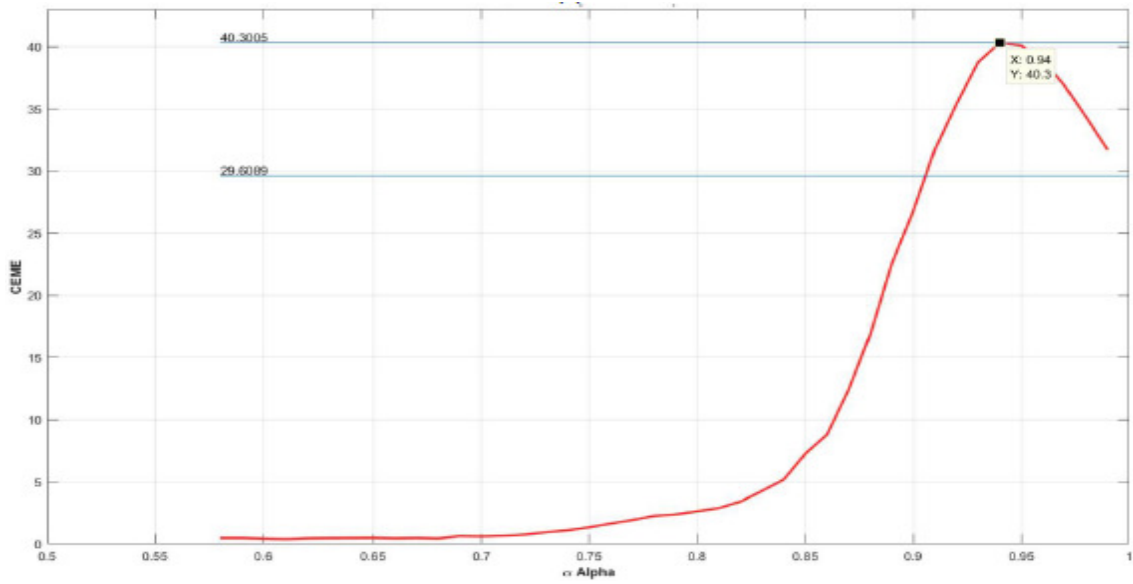

(b)

Figure 3: (a) Original Image "football.jpg"[31]; (b) Plot of CEME vs alpha for alpha-rooting method by 2-D QDFT.

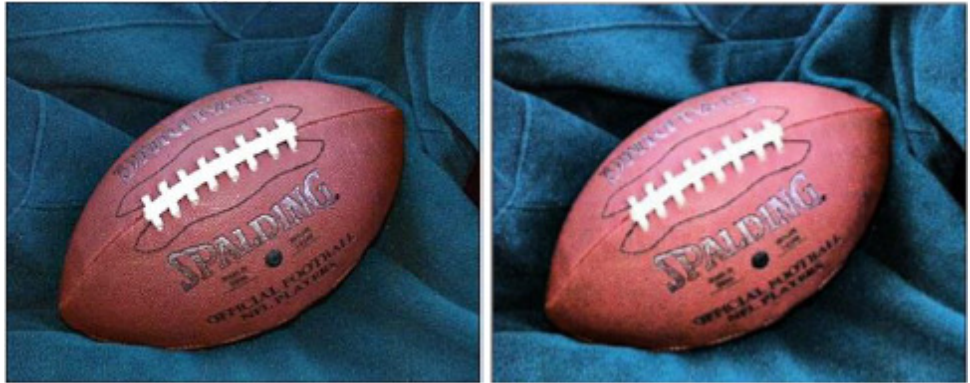

(c)  (d)

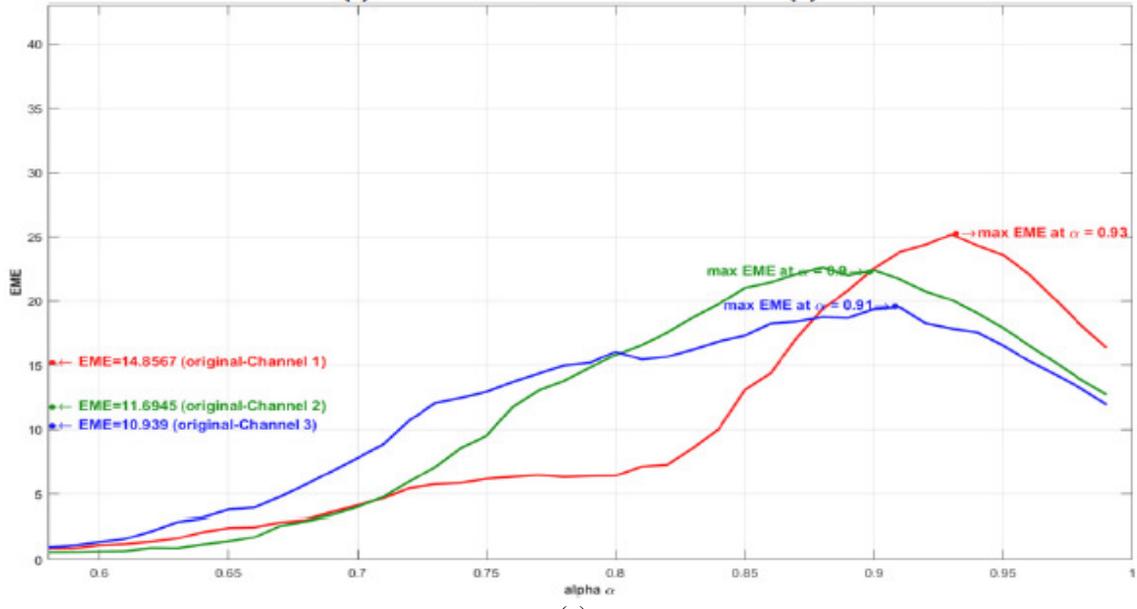

(e)

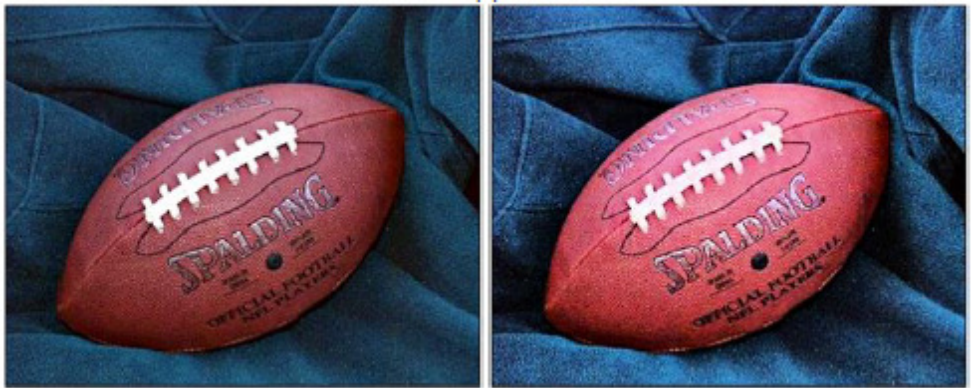

(f)  (g)

Figure 3: (continued) (c) Enhanced image by Alpha-rooting by 2-D QDFT; (d) Histogram equalised image of the image in (c); (e) Plot of EME vs alpha for alpha-rooting by 2-D DFT for each channel (R,G,B), (f) Enhanced image by alpha-rooting by 2-D DFT; (g) Histogram equalised image of image in (f).

Figure 3 shows the enhancement image results of the original image "football.jpg" in Fig. 3(a). The plot of CEME vs alpha in Fig. 3(b) gives the best value of alpha at 0.94. The CEME of the enhanced image by the 2-D QDFT alpha-rooting method has maximum CEME at alpha equals 0.94. The resulting enhanced image by the method of alpha-rooting by the 2-D QDFT is shown in Fig. 3(c) and its enhancement by histogram equalization is shown in Fig. 3(d). From the plot of EME vs alpha for the image "football.jpg" in Fig. 3(e), we see that the best value of alpha for channels red, green and blue are respectively 0.93, 0.9, and 0.91. With the best alpha value, each channel is enhanced separately by the alpha-rooting by the 2-D DFT and the resulting image is further enhanced spatially by histogram equalization. The enhanced image are respectively shown in parts (f) and (g) of Fig. 3.

The enlisted values in Table 3 also show the validity of the statement that the enhancement of the image "football.jpg" gives better enhancement image results when enhanced by the method of alpha-rooting by the 2-D QDFT than when enhanced by the method of alpha-rooting by the 2-D DFT. The further spatial transformation by the histogram equalization gives higher CEME value for the image pre-processed by the alpha-rooting by the 2-D QDFT than when pre-processed by the channel-by- channel approach of alpha-rooting by the 2-D DFT.

Table 3: Alpha and CEME/EME values of Original and Enhanced Image of "football.jpg".

| "football.jpg" | CEME | Alpha | EME |
|---|---|---|---|
| Original Image | 25.6089 | | R: 14.8567 |
| | | | G: 11.6945 |
| | | | B: 10.9390 |
| 2-D QDFT Alpha-Rooting | 40.3005 | 0.94 | - |
| 2-D QDFT Alpha-Rooting With Spatial Transformation | 45.1842 | 0.94 | - |
| 2-D DFT Alpha-Rooting | 37.2225 | R: 0.93 | 25.1773 |
| | | G: 0.90 | 22.3891 |
| | | B: 0.91 | 19.5084 |
| 2-D DFT Alpha-Rooting with Spatial Transformation | 40.9300 | R: 0.93 | |
| | | G: 0.90 | |
| | | B: 0.91 | |

Fig. 4 shows the original image "autumn.tif" in Fig. 4(a) enhanced in the frequency domain by alpha-rooting by the 2-D QDFT in Fig. 4(c) and further enhanced spatially by the histogram equalization in Fig. 4(d). The best value of alpha is obtained from the plot of CEME vs alpha in Fig. 4(b). The alpha chosen for the enhancement algorithm is 0.95. The image "autumn.tif" is also enhanced by the channel-by-channel approach of enhancement by alpha-rooting by the 2-D DFT. The resulting image and the image after spatially enhancing by histogram equalization is shown parts (f) and (g) respectively of Fig. 4. The alpha values chosen are the same and equal 0.94 for all the channel.

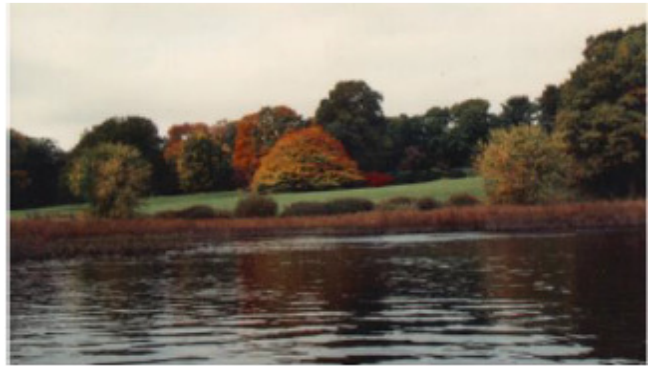

(a)

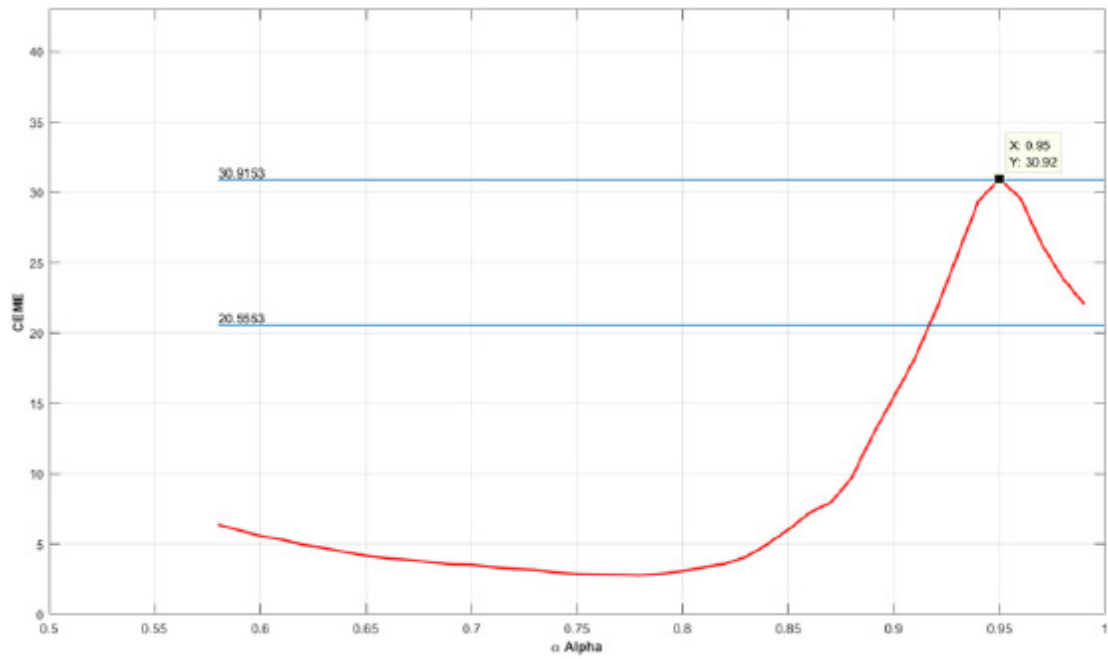

(b)

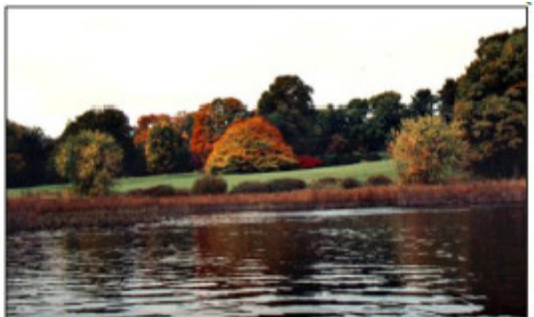 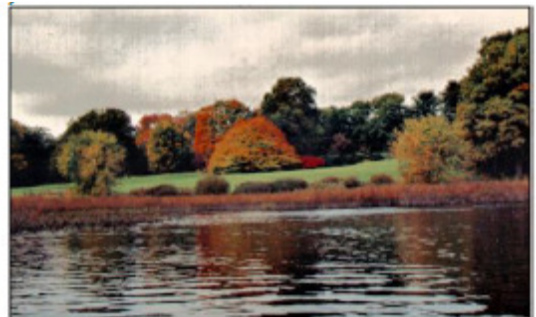

(c) (d)

Figure 4: (a) Original Image "autumn.tif"[33], (b) Plot of CEME vs alpha for Alpha-rooting Method by 2-D QDFT; (c) Enhanced image by alpha-rooting by 2-D QDFT; (d) Histogram equalised image of the image in (c);

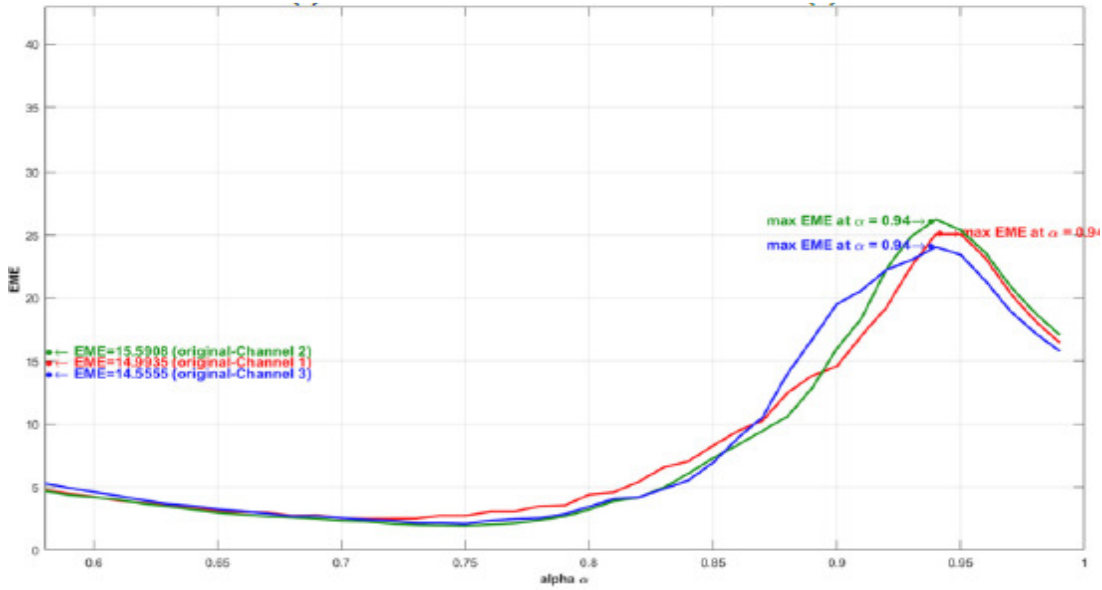

(e)

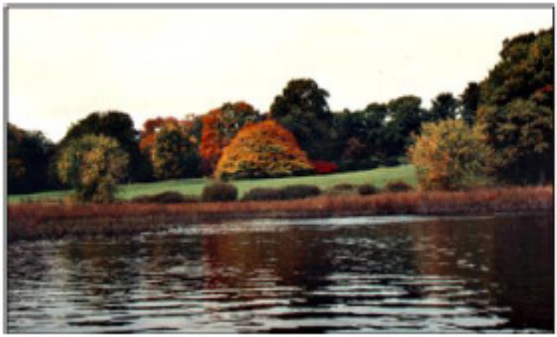 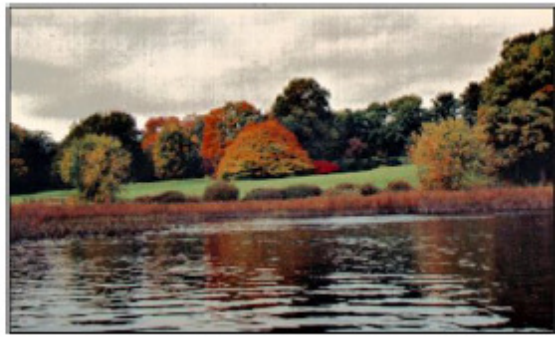

(f) (g)

Figure 4: (continued) (e) Plot of EME vs alpha for Alpha-rooting by 2-D DFT for each channel (R,G,B); (f) Enhanced image by alpha-rooting by the 2-D DFT; (g) Histogram equalised image of image in (f).

Table 4 shows the comparative results of enhancement by both methods. We can see that the enhancement measure CEME is greater in both cases of enhancement algorithm based on quaternion approach than the CEME values of enhanced image based on the channel-by-channel approach of the alpha-rooting method by the 2-D DFT followed by the histogram equalization.

Fig. 5 shows the enhancement image results of the original image "image8.jpg" shown in Fig. 5(a). This is a very good example of the advantage of the proposed algorithm of the alpha-rooting method by the 2-D QDFT followed by spatial transformation. The quaternion approach of alpha-rooting method gives a higher value of CEME value as compared with the channel-by-channel approach of the alpha-rooting method.

Table 4: Alpha and CEME/EME values of Original and Enhanced Image of "autumn.tif".

| "autumn.tif" | CEME | Alpha | EME |
|---|---|---|---|
| Original Image | 20.5553 | | R: 14.9935 |
| | | | G: 15.5908 |
| | | | B: 14.5555 |
| 2-D QDFT Alpha-Rooting | 30.9153 | 0.95 | - |
| 2-D QDFT Alpha-Rooting With Spatial Transformation | 37.4616 | 0.95 | - |
| 2-D DFT Alpha-Rooting | 29.7669 | R: 0.94 | 25.0809 |
| | | G: 0.94 | 26.2241 |
| | | B: 0.94 | 24.0506 |
| 2-D DFT Alpha-Rooting with Spatial Transformation | 34.0251 | R: 0.94 | |
| | | G: 0.94 | |
| | | B: 0.94 | |

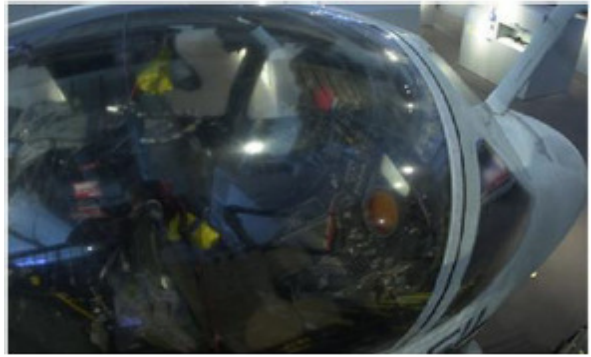

(a)

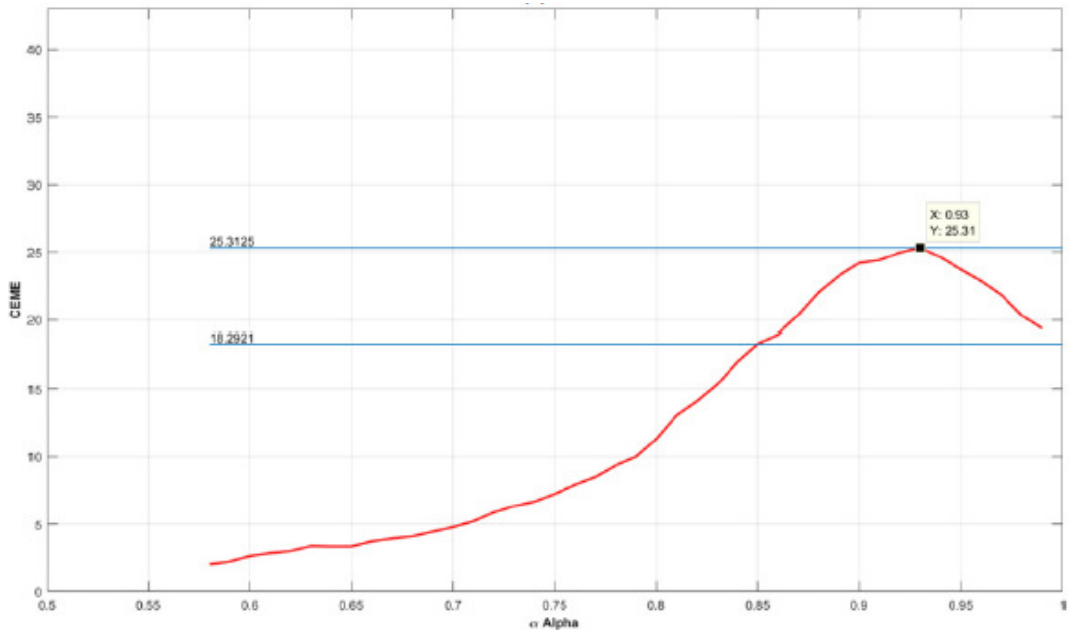

(b)

Figure 5 (a) Original Image "image8.jpg"[32]; (b) Plot of CEME vs alpha for Alpha-rooting Method by 2-D QDFT;

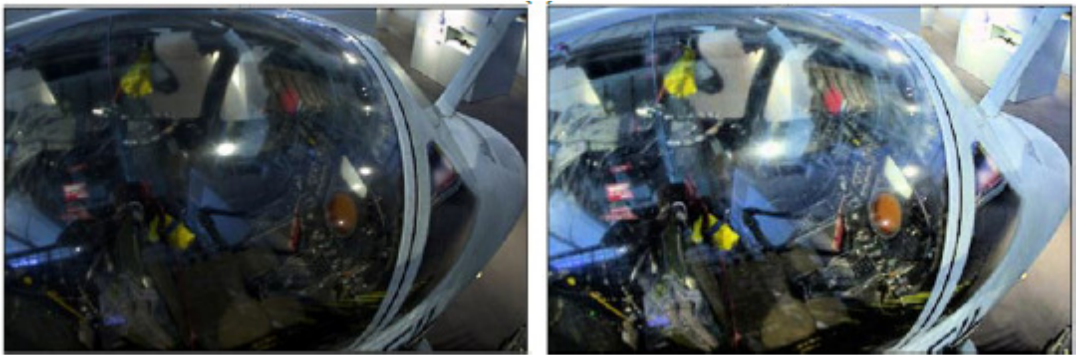

(c)              (d)

Figure 5 (continued) (c) Enhanced image by alpha-rooting by 2-D QDFT (d) Histogram equalised image of the image in (c).

The alpha-rooting method by the 2-D QDFT enhances the image in the frequency domain. The low frequency and high frequency of the image are enhanced by the alpha-rooting method by the 2-D QDFT, as shown in Fig. 5(c). The further enhancement in the spatial domain enhances the image result further. We see in Fig. 5(d) that additional spatial transformation of the enhanced image in Fig. 5(c) brightens the image and shows more details of the image information. The best alpha value is chosen from the plot in Fig. 5(b) and the chosen alpha is 0.93 for the alpha-rooting method by the 2-D QDFT.

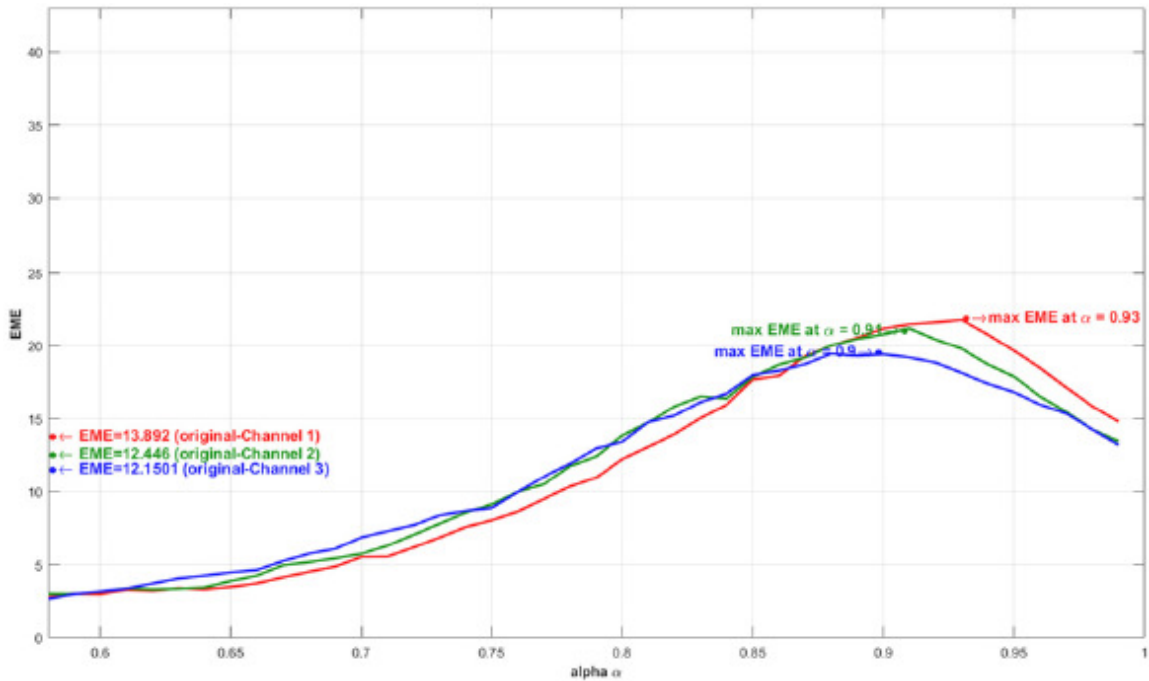

(e)

Figure 5: (continued) (e) Plot of EME vs alpha for Alpha-rooting by 2-D DFT for each channel (R,G,B);

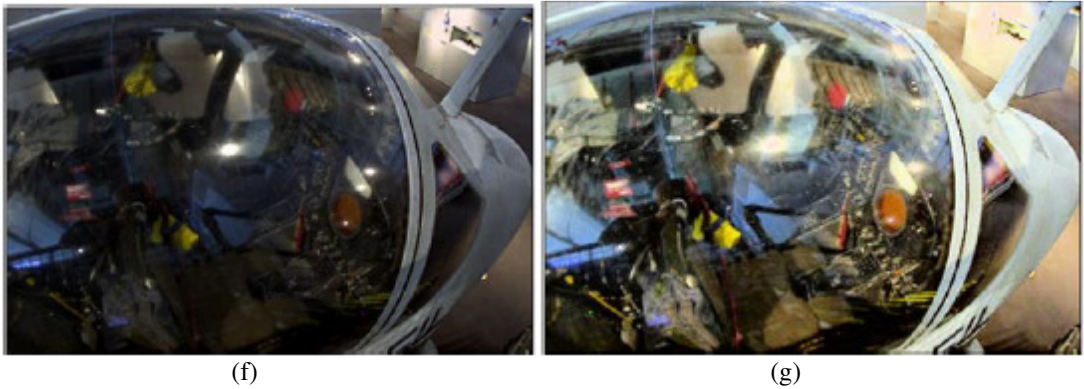

(f)            (g)

Figure 5: (continued) (f) Enhanced image by alpha-rooting by the 2-D DFT; (g) Histogram equalized image of image in (f).

The three alpha values needed for the algorithm of the alpha-rooting method of 2-D DFT are chosen from the plot of EME vs alpha in Fig. 5(e). Since this approach is channel-by-channel enhancement, each channel is enhanced separately and the resulting enhanced channels are used to compose the resulting enhanced image. The three alpha values chosen are 0.93, 0.91, and 0.9 for the red, green, and blue respectively.

Table 5 gives comparative results of the enhancement of image "image8.jpg" by the two methods. The image enhanced by the quaternion approach of the alpha-rooting method by the 2-D QDFT followed by spatial transformation by histogram equalization gives better CEME values as compared to the corresponding enhanced image by the channel-by-channel approach of alpha-rooting by the 2-D DFT followed by the histogram equalization.

Table 5: Alpha and CEME/EME values of Original and Enhanced Image of "image8.jpg".

| "image8.jpg" | CEME | Alpha | EME |
|---|---|---|---|
| Original Image | 18.2921 | | R: 13.8920 |
| | | | G: 12.4460 |
| | | | B: 12.1501 |
| 2-D QDFT Alpha-Rooting | 25.3125 | 0.93 | - |
| 2-D QDFT Alpha-Rooting With Spatial Transformation | 29.5053 | 0.93 | - |
| 2-D DFT Alpha-Rooting | 23.5409 | R: 0.93 | 21.7847 |
| | | G: 0.91 | 21.2000 |
| | | B: 0.90 | 19.4170 |
| 2-D DFT Alpha-Rooting with Spatial Transformation | 25.0866 | R: 0.93 | |
| | | G: 0.91 | |
| | | B: 0.90 | |

## 4. CONCLUSIONS

The enhancement of color images by the 2-D QDFT based alpha-rooting followed by spatial transformations like the histogram equalization show good enhancement results. The quaternion approach gives a color to the image that is closer in the original color of the image. That is, the enhanced image keeps a better harmony in the combinations of color components as compared with the images when processed individually. The CEME measure is higher in quaternion approach of image enhancement than the channel-by-channel approach of image enhancement. In the future work, psychophysics of the visual perception of color in the images will be studied in detail, to understand better the effects of the color.

## AUTHORS

**Artyom M. Grigoryan** received the PhD degree in Mathematics and Physics from Yerevan State University (1990). He is an associate Professor of the Department of Electrical Engineering in the College of Engineering, University of Texas at San Antonio. The author of four books, 9 book-chapters, 3 patents and more than 120 papers and specializing in the design of robust filters, fast transforms, tensor and paired transforms, discrete tomography, quaternion imaging, image encryption, processing biomedical images.

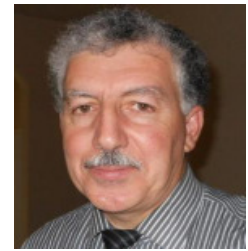

**Aparna John** received her B. Tech. degree in Applied Electronics and Instrumentation from University of Calicut, India and M. Tech. degree in Electronics and Communication with specialization in Optoelectronics and Optical Communication from University of Kerala, India. Then she joined as lecturer in the Department of Optoelectronics, University of Kerala, India, for a short-term period, from 2008 to 2011. Now, she is a doctoral student in Electrical Engineering at University of Texas at San Antonio. She is pursuing her research under the supervision of Dr. Artyom M. Grigoryan. Her research interests include image processing, color image enhancements, fast algorithms, quaternion algebra and quaternion transforms including quaternion Fourier transforms. She is a student member of IEEE and also a member of Eta Kappa Nu Honor Society, University of Texas San Antonio

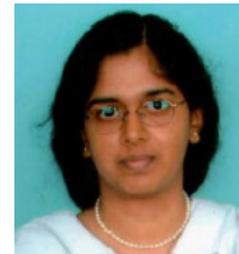

**Sos S. Agaian** is the Distinguished Professor at the City University of New York/CSI. Dr. Agaian received the Ph.D. degree in math and physics from the Steklov Institute of Mathematics, Russian Academy of Sciences, and the Doctor of Engineering Sciences degree from the Institute of the Control System, Russian Academy of Sciences. He has authored more than 500 scientific papers, 7 books, and holds 14 patents. His research interests are Multimedia Processing, Imaging Systems, Information Security, Artificial Intelligent, Computer Vision, 3D Imaging Sensors, Fusion, Biomedical and Health Informatics

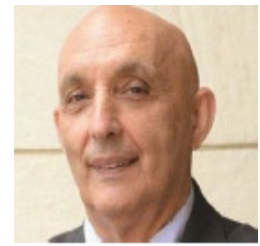